\documentclass[twocolumn,aps,prc,superscriptaddress,showpacs]{revtex4}
\usepackage{amsmath, amsthm, amssymb, bm, braket}
\usepackage{epsfig,graphics}
\usepackage{graphicx}% Include figure files
\usepackage{color}
\usepackage{fancybox}
\usepackage{bm}% bold math
\begin{document}

%\begin{CJK*}{GBK}{song}

\title{The role of the Heisenberg principle in Constrained Molecular Dynamics model}

\author{K. Wang}
\email[]{wangkang@sinap.ac.cn}
\affiliation{Shanghai Institute of Applied Physics, Chinese Academy of Sciences, Shanghai 201800, China}
\affiliation{University of the Chinese Academy of Sciences, Beijing 100080, China}
\affiliation{School of Physical Science and Technology, ShanghaiTech University, Shanghai 201203, China}

\author{A. Bonasera}
\affiliation{Cyclotron Institute, Texas A\&M University, College Station, TX 77843, USA}
\affiliation{Laboratori Nazionali del Sud, INFN, via Santa Sofia, 62, 95123 Catania, Italy}

\author{H. Zheng}
\affiliation{School of Physics and Information Technology, Shaanxi Normal University, Xi'an 710119, Shaanxi, China}

\author{G. Q. Zhang}
\affiliation{Shanghai Institute of Applied Physics, Chinese Academy of Sciences, Shanghai 201800, China}

\author{Y. G. Ma}
\affiliation{Shanghai Institute of Applied Physics, Chinese Academy of Sciences, Shanghai 201800, China}
\affiliation{School of Physical Science and Technology, ShanghaiTech University, Shanghai 201203, China}
\affiliation{Institute of Modern Physics, Department of Nuclear Science and Technology, Fudan University, Shanghai 200433, P.R. China}

\author{W. Q. Shen}
\affiliation{Shanghai Institute of Applied Physics, Chinese Academy of Sciences, Shanghai 201800, China}
\affiliation{School of Physical Science and Technology, ShanghaiTech University, Shanghai 201203, China}

\date{\today}

%%%%%%%%%%%%%%%%%%%%%%%%%%%%%%%%%%%%%%%%%%%%%%%%%%%%%%%%%

\renewcommand{\figurename}{FIG.}

\def \beq{\begin{equation}}
\def \eeq{\end{equation}}
\def \beqa{\begin{eqnarray}}
\def \eeqa{\end{eqnarray}}

\begin{abstract}
    We implement the Heisenberg principle into the Constrained Molecular Dynamics (CoMD) model with a similar approach to the Pauli principle using the one-body occupation probability $\bar{f}_i$. Results of the modified and the original model with comparisons to data are given. The binding energies and the radii of light nuclei obtained with the modified model are more consistent to the experimental data than the original one. The collision term and the density distribution are tested through a comparison to p+$^{12}$C elastic scattering data. Some simulations for fragmentation and superheavy nuclei production are also discussed.

\end{abstract}

\pacs{21.60.Ka, 24.10.Lx, 25.70.-z, 25.70.Jj}
\maketitle

\section{Introduction} \label{Sec:intro}

    In recent decades, heavy-ion reactions have been an important approach to study the properties of nuclear structure and reaction dynamics, such as the properties of exotic nuclei, the nuclear equation of state, and superheavy nuclei among other features. Besides experiments, theoretical models become more and more useful in these researches. Among them, the molecular dynamics models achieved great success since they dealt better with the $N$-body problem than the one-body semiclassical transport models like the Vlasov-Uehling-Uhlenbeck (VUU)~\cite{bertsch1988} and the Boltzmann-Nordheim-Vlasov (BNV) models~\cite{bonasera1994,papa2001}. The Fermionic Molecular Dynamics model(FMD)~\cite{feldmeier1990,feldmeier1995} and the Antisymmetrized Molecular Dynamics model(AMD)~\cite{ono1992a,ono1992b} have successfully preserved the Fermionic nature of the many body system by expressing the wave function of the system as a single Slater determinant of $N$ wave packets. As a result, the numerical effort of these two models grows with $N^{3-4}$. It takes very long CPU time for calculations of large mass systems ($\ge$80 nucleons) so that approximations are needed~\cite{feldmeier2000}. The Quantum Molecular Dynamics model (QMD)~\cite{aichelin1986,aichelin1991}, which needs only double-fold loops to calculate two-body interactions, making its numerical effort grow with $N^2$~\cite{papa2001}, is much faster than FMD and AMD. It has made great achievements in heavy-ion reactions from intermediate to high energies~\cite{hartnack1998,ma2006,kumar2010,gautam2012,feng2018}. However, the Fermionic feature of nucleons is lacking in the QMD model, especially in the ground states or in low-energy reaction phenomena~\cite{papa2001}. Several modified quantum molecular dynamics models have been proposed to solve this problem, like the extended Quantum Molecular Dynamics (EQMD) model~\cite{maruyama1996}, and the Constrained Molecular Dynamics (CoMD) model~\cite{papa2001,papa2005}.

    The EQMD model tried to mimic the Fermionic features by introducing the Pauli potential, which forbids nucleons with the same spin and isospin from coming close to each other in phase space~\cite{maruyama1996}. The model shows some good features such as stable ground states and saturation properties of nuclear matter, but it also has some problems, for instance a spurious repulsion in the collisions~\cite{wada1998, he2014, papa2001}.

    By constraining the phase space to fulfill the Pauli principle at each time step, the CoMD model has successfully introduced the effect of the Fermionic nature to the nuclear many-body system. It is able to reproduce with the same set of parameters not only the main characteristic of stable nuclei in a wide mass region ($A$=30-208) but also the experimental fragment charge distribution in many collisions such as $^{40}$Ca+$^{40}$Ca, $^{197}$Au+$^{197}$Au, $^{112,123}$Sn+$^{58,64}$Ni systems at 35 MeV/nucleon, and $^{40}$Ca+$^{48}$Ca system at 25 MeV/nucleon~\cite{papa2001, papa2003}. However, the binding energies and theoretical prediction of light nuclei, in particular the $Z$=2 yield, are not consistent with experimental results~\cite{papa2001}. In this paper, we try to improve these features by introducing the Heisenberg principle as another constraint into the CoMD model and study its effect on the finite nuclei. Similar mechanism has already been successfully applied to study the penetrability of the Coulomb barrier with molecular dynamics simulations~\cite{kimura2004}, and also the atomic structure and collision calculations~\cite{kimura2005}.

    The paper is organized as follows. Section \ref{Sec:formalism} gives a brief introduction of the CoMD model as well as the formalism to implement the Heisenberg principle into the original model. Section \ref{Sec:res} presents the results and discussions of the model. The binding energies of nuclei during initialization and time evolution from the modified and the original models and the comparisons to data are given. A brief summary is given in Sec. \ref{Sec:sum}.

\section{Model and Formalism} \label{Sec:formalism}

    The nucleon wave functions and the mean field used in this work are the same as in the original CoMD model to maintain the same nuclear equation of state. Details about the effective interaction were described in Ref.~\cite{papa2001}.

    In the original CoMD model, the constraints for the Pauli principle at each time step are achieved via the following:
        \beqa
            {\bar{f}_i}
            &\le& 1, \label{eq:fd}\\
            \bar{f}_i
            &\equiv& \sum_{j}\delta_{\tau_i,\tau_j}\delta_{s_i,s_j}\int_{h^3}f_j({\it r,p})d^3rd^3p, \label{fi}
        \eeqa
    in which $\tau_i$ represents the nucleon isospin and $s_i$ refers to the spin projection quantum number~\cite{papa2001}. At each time step and for each particle $i$, an ensemble $K_i$ of nearest identical particles (including the particle $i$) is determined within the distances $\sqrt{\pi\hbar\sigma{_r}/2\sigma{_p}}$ and $\sqrt{\pi\hbar\sigma{_p}/2\sigma{_r}}$ from particle $i$ in $r$ and $p$ spaces, respectively~\cite{papa2001}. Here $\sigma_r$ represents the width of the Gaussian wave function in $r$ space and $\sigma_p$ is taken as the width in $p$ space. Then the phase space occupation $\bar{f}_i$ is calculated for the identical particles. If $\bar{f}_i$ is greater than 1, we change randomly the momenta of the particles belonging to the ensemble $K_i$ to make $\bar{f}_i$ less than 1. The mechanism we used is similar to a many-body elastic scattering so that the total momentum and the total kinetic energy of the newly generated sample are conserved. The new sample is accepted only if the phase-space occupation $\bar{f}_i$ is smaller than the one before~\cite{papa2001}.

    We adopt a similar approach to introduce the Heisenberg principle into the modified model. After the check of the Pauli principle for identical particles (same spin and isospin) as mentioned before, for each of those particles that do not violate the Pauli principle, we search the ensemble $L_i$ of the nearest non-identical particles (including the particle $i$) within the distances $\sqrt{\pi\hbar\sigma{_r}/2\sigma{_p}}$ and $\sqrt{\pi\hbar\sigma{_p}/2\sigma{_r}}$ from the particle $i$ in $r$ and $p$ spaces, respectively~\cite{papa2001}. Then we calculate the average $\bar{f}_i$ for the non-identical particles in the ensemble $L_i$. If the average of $\bar{f}_i$ has a value greater than 25.0/93.0 (a value obtained by calculating the $\bar{f}_i$ of two nucleons with their $\Delta\it{r}\Delta\it{p}$ equals to $\hbar/2$), we change the momenta of the particles belonging to the ensemble $L_i$ as we did for $K_i$ before. Notably, those particles which have been scattered because of the Pauli principle will not be included in the check of the Heisenberg principle. This ensures that one particle will not be scattered twice.

    Similar to the Pauli blocking, we introduced the "Heisenberg blocking" in the collision term in the modified model. For each $NN$ collision, we calculate the occupation probability as we did before and check for both the Pauli principle and the Heisenberg principle. If they are both satisfied, the collision is accepted, rejected otherwise. In this way, the Heisenberg principle is important for light nuclei ($Z\le4$) and the Pauli principle becomes dominant for heavy nuclei.

    We also modified the cooling procedure in the initialization step to prepare the ground state of the nucleus in the modified model. We use the constraints to solve the equations of motion of the nucleons with friction terms. The nucleons are first distributed in spheres of radius 1.2${\times}A^{1/3}$ fm and $P^{nm}_F$ (Fermi momentum for infinite nuclear matter) in coordinate and momentum space, respectively. Then at each time step, we first calculate the value of $\bar{f}_i$ to fulfill the Pauli principle. If $\bar{f}_i$ is greater than 1, the momenta of the particles belonging to the ensemble $K_i$ are scaled by a factor of 1.005. If $\bar{f}_i$ is less than 1, we calculate the value of $\bar{f}_i$ to fulfill the Heisenberg principle for the particles belonging to ensemble $L_i$. If the average $\bar{f}_i$ for the Heisenberg principle is greater than 25.0/93.0, we set the scale factor to 1.0005. If this $\bar{f}_i$ is also less than the threshold value, the scale factor is set to 0.99. Roughly, by iterating this cooling procedure up to 1000 fm/c, we can get stationary values of the total binding energy and the radius of the nucleus for which the ground state is wanted for the heavy-ion reaction simulation or other goals. After this cooling procedure, the friction term is switched off and we test the stability of the ground state by checking the time evolution of its root mean square radius $R$. The prepared ground state will be accepted only if $R$ is stable for at least 1000 fm/c.

\section{Results and Discussions} \label{Sec:res}

    In order to examine the CoMD model with the Heisenberg principle (CoMD$_H$), we compare some simulated results from CoMD$_H$ to the ones with the same conditions from the original model and also to the experimental data.

    \begin{figure}[htbp] \begin{center}
        \includegraphics[width=\hsize]{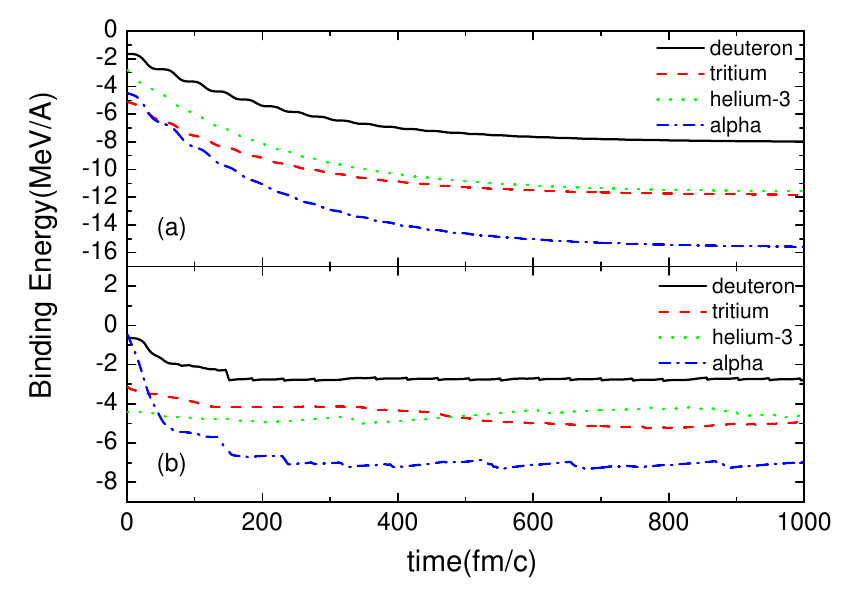}
        \caption{(Color online) Time evolution of the binding energies during the cooling step for the original model without the Heisenberg principle (a) and the modified model with the Heisenberg principle (b).}
    \label{fig:te}
    \end{center} \end{figure}

\begin{figure}[htbp] \begin{center}
        \includegraphics[width=\hsize]{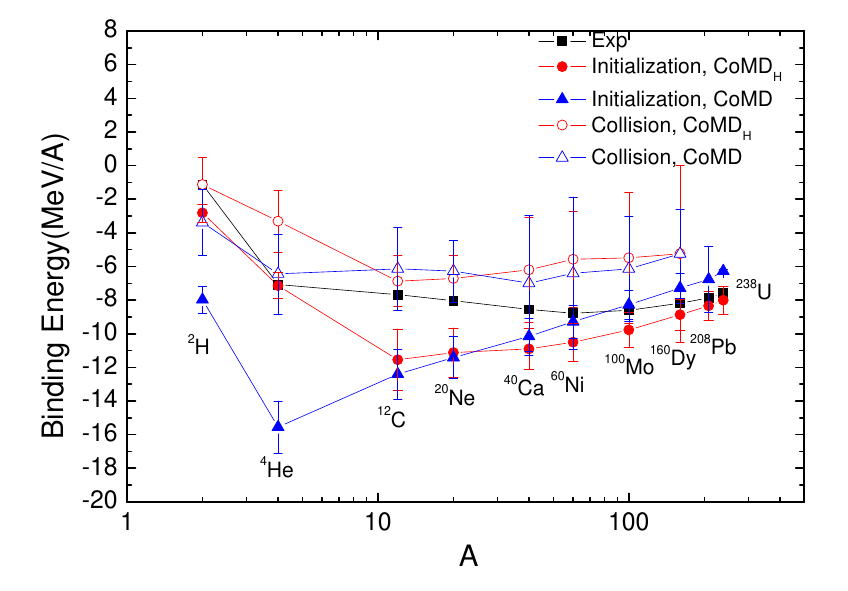}
        \caption{(Color online) Binding energies of nuclei with mass ranging from 2 to 238 in initializations (full symbols) and collisions (open symbols). The (black) full squares represent the experimental data. The (red) solid circles and the (blue) solid triangles represent the binding energies of nuclei during the initialization calculated with the modified model including the Heisenberg principle (CoMD$_H$) and with the original model (CoMD), respectively. The (red) open circles and the (blue) open triangles represent the binding energies of the fragments obtained in the nucleus-nucleus collisions. The latter fragments might be stable but not in their ground state because they would normally deexcite say by $\gamma$ rays at very long times, a mechanism not included in the model.}
    \label{fig:bech}
    \end{center} \end{figure}

    In the original model, the binding energies of the light particles are not satisfactory, similarly to other approaches where the constant distribution width is added to the binding energy~\cite{feldmeier1990,ono1992a,aichelin1986,aichelin1991,giuliani2014}. Figure \ref{fig:te} shows the comparisons of the binding energies of the light nuclei calculated with the original and the modified models. From the results of the original model, we see that the binding energies of the light nuclei keep going down to a very low value and they are over bound. This is due to the fact that these light particles, i.e.,  deuteron, tritium, helium-3, and alpha particles do not violate the Pauli principle thus their gaussian wave packets can move on top of each other in the cooling procedure. We stress that this feature is common to all molecular dynamics models but other authors choose to include the gaussian finite width as the kinetic energy part which does not affect the motion of nucleons at all. Thus their ground states are solids~\cite{feldmeier2000,papa2005,giuliani2014}. When adding the Heisenberg principle, we see that the binding energies of the light nuclei oscillate around some average values. These are the real ground state of the model depending only on the interaction parameters and especially the surface term for light nuclei. The binding energies of these light nuclei are now closer to the experimental values even though deuterium, tritium, and $^3$He are still over bound. Figure \ref{fig:bech} shows the binding energies of some selected nuclei (the mass is ranging from 2 to 238) from CoMD and CoMD$_H$ calculations. The experimental data are retrieved from the National Nuclear Data Center~\cite{pritychenko2012}. The modified model seems to work better throughout the whole nuclei chart than the original one. We stress also that in order to fulfill the Heisenberg principle, we enforce its effect in the collision term as well for non-identical particles.

\begin{figure}[htbp] \begin{center}
        \includegraphics[width=\hsize]{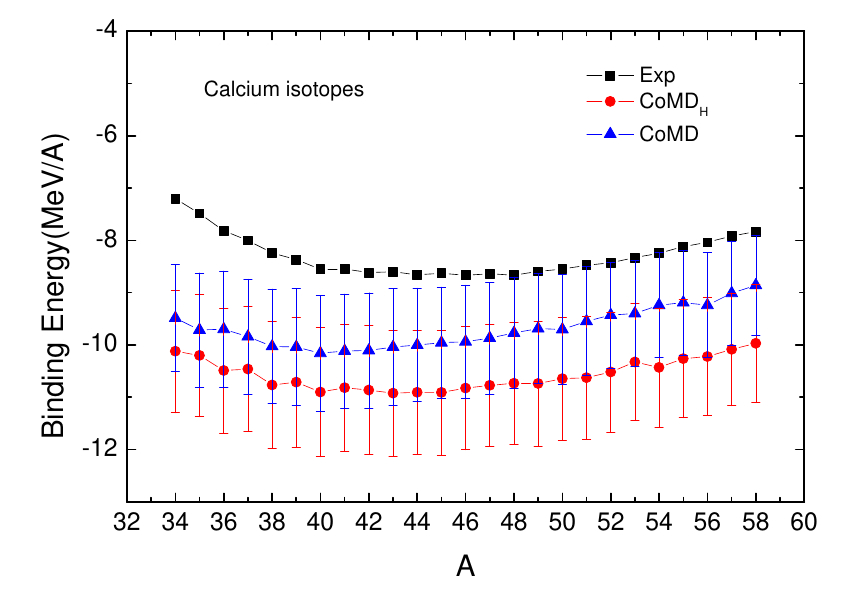}
        \caption{(Color online) Binding energies of the calcium isotopes.}
    \label{fig:beca}
    \end{center} \end{figure}

\begin{figure}[htbp] \begin{center}
        \includegraphics[width=\hsize]{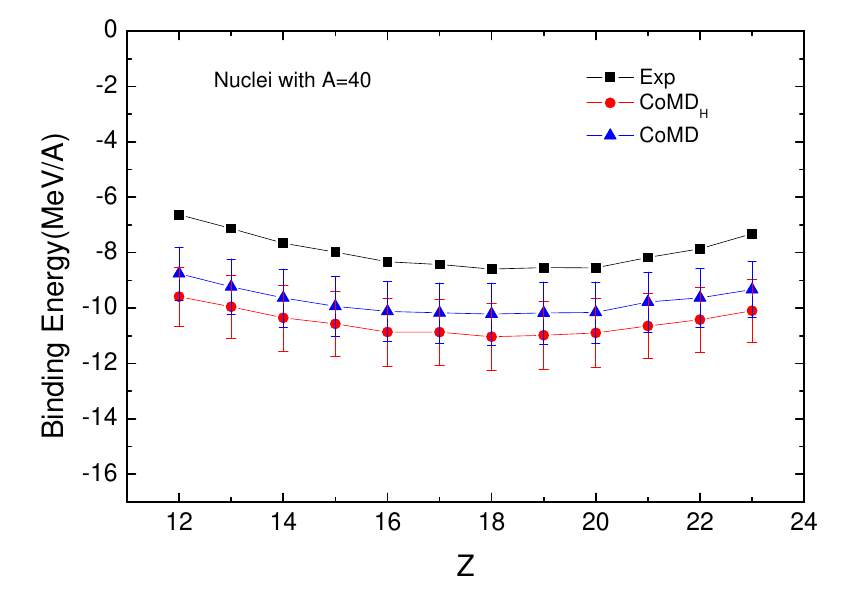}
        \caption{(Color online) Binding energies of nuclei with mass $A=40$.}
    \label{fig:bese}
    \end{center} \end{figure}

    To test the ability of the modified model to work with exotic nuclei, we calculated the binding energies of the ground state of whole calcium isotopes as shown in Fig. \ref{fig:beca}. Though the binding energies calculated by the modified model are a little more bound, little differences (within the calculations uncertainty) to the original approach are observed. Figure \ref{fig:bese} shows the binding energies of the nuclei with mass $A=40$. Both models reproduce the binding energies in a similar fashion within the error bars, thus suggesting that with larger nuclei the Heisenberg principle is less and less important than the Pauli one as expected. Clearly the reproduction of the data is not perfect and we did not try to adjust parameters since we believe that important effects, such as (iso)spin-(iso)spin forces are missing and we plan to include those effects in future works.

\begin{figure}[htbp] \begin{center}
        \includegraphics[width=\hsize]{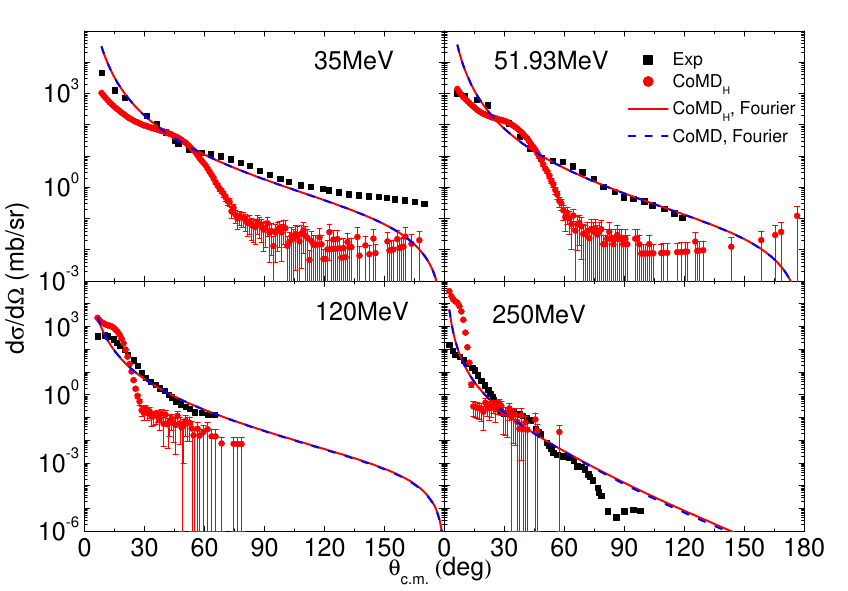}
        \caption{(Color online) Comparison of calculated cross sections for the elastic scattering of 35 MeV, 51.93 MeV, 120 MeV, and 250 MeV protons from $^{12}$C with data~\cite{pignanelli1986, ohnuma1980, comfort1981, meyer1988, kim2008}. The (black) full squares represent the experimental data. The (red) solid circles represent the calculated results with the CoMD$_H$ model. The (red) solid and the (blue) dashed lines represent the cross sections calculated with the Fourier transform of the charge density distribution (see the discussion in the context) for the CoMD$_H$ and the CoMD models, respectively.}
    \label{fig:sc}
    \end{center} \end{figure}

     In order to test the reliability of the modified collision term coupled to the mean field, we performed elastic scattering simulations for the p+$^{12}$C system at different energies. The collisions were performed in the range of impact parameters from 0 to 6 fm. Since the proton is scattered from the target without losing energy (we included a little energy loss below 4 MeV, the first excited level of $^{12}$C), the reaction is rather fast and we followed its time evolution up to 200 fm/c. In Fig. \ref{fig:sc}, the cross sections for the elastic scattering data of 35 MeV~\cite{pignanelli1986}, 51.93 MeV~\cite{ohnuma1980}, 120 MeV~\cite{comfort1981}, and 250 MeV~\cite{meyer1988} protons from $^{12}$C are compared with the results of our model. As shown in the figure, the results of the modified model are in good agreement with the experimental data except for the large angle parts, mostly due to low statistics (1,000,000 events for each case). Because of the large numerical effort, we performed calculations for the modified model only. Inelastic scattering calculations required a much longer CPU time and we decided to postpone them until (iso)spin-(iso)spin forces are included in the model. Pion production is not included in the model, but its effect might become important at the highest beam energy. However, since we are looking at the elastic scattering data only, we feel our calculations are reliable.

\begin{figure}[htbp] \begin{center}
        \includegraphics[width=\hsize]{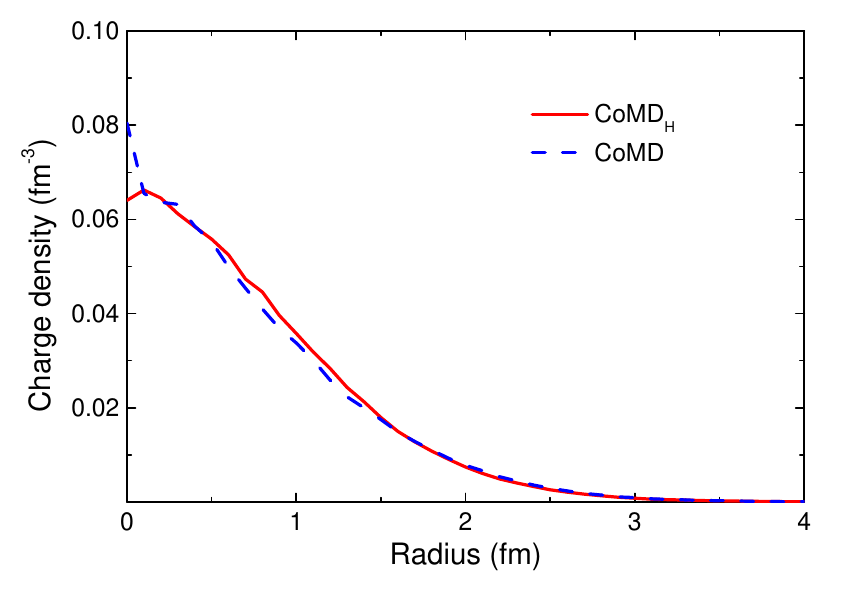}
        \caption{(Color online) Comparison of calculated density distributions in the ground state of $^{12}$C.}
    \label{fig:dc}
    \end{center} \end{figure}

     We also adopted another method usually used in the literature to test our results~\cite{preston1975}. First we calculated the charge density distribution in the ground state of  $^{12}$C displayed in Fig. \ref{fig:dc}. The Fourier transform of the charge density times the Mott cross section~\cite{preston1975} gives the elastic cross sections reported in Fig. \ref{fig:sc} as full lines (CoMD$_H$) and dashed lines (CoMD). The small difference in the densities from the two models does not produce any appreciable difference in the elastic scattering. The agreement with data is satisfactory and seems to agree better with increasing proton beam energy, a similar feature noticed in the numerical simulations.

\begin{figure}[htbp] \begin{center}
        \includegraphics[width=\hsize]{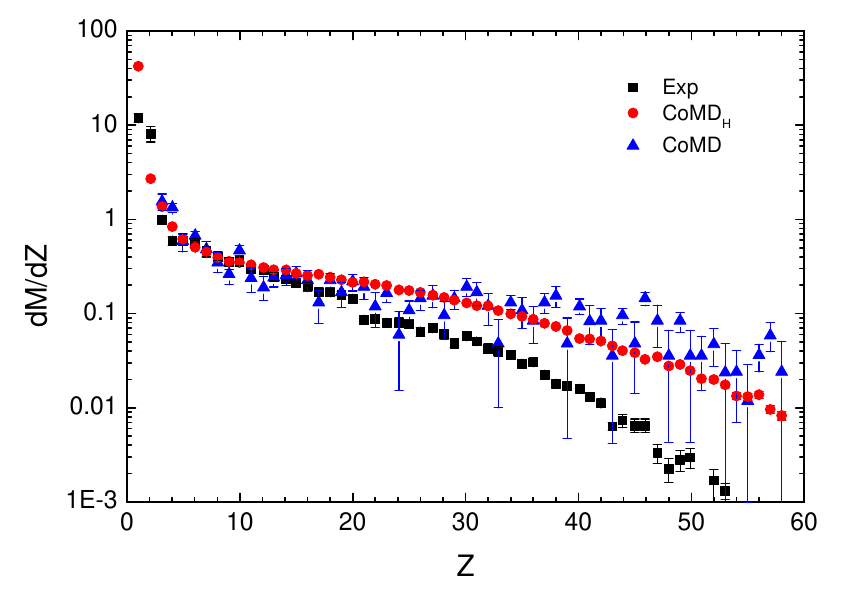}
        \caption{(Color online) Comparison between the experimental isotope distribution and the calculated results of the CoMD$_H$ and the CoMD models for $^{197}$Au+$^{197}$Au at 35 MeV/A with impact parameters ranging from 0 to 3.5 fm and t=1500 fm/c~\cite{desequelles1998, papa2001}.}
    \label{fig:aa}
    \end{center} \end{figure}

\begin{figure}[htbp] \begin{center}
        \includegraphics[width=\hsize]{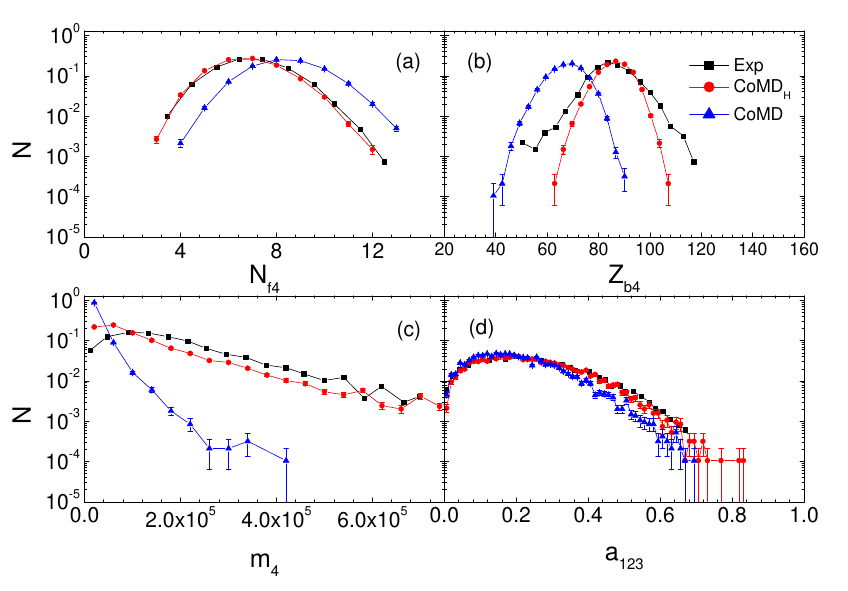}
        \caption{(Color online) Comparison of the observable distributions for the experimental data (black full squares), the CoMD$_H$ (red solid circles) and the CoMD (blue solid triangles) calculations for the $^{197}$Au+$^{197}$Au at 35 MeV/A with impact parameters ranging from 0 to 3.5 fm and t=1500 fm/c~\cite{desequelles1998, papa2001}. (a) Fragment (Z $\ge$ 4) multiplicity ($N_{f4}$), (b) Total charge bound in form of fragments ($Z_{b4}$), (c) Fourth moment $m_4$ of the fragment charge distribution with the heaviest fragment excluded, (d) Charge asymmetry distribution. All the vertical axes represent the mean multiplicity per event of the considered observable~\cite{desequelles1998}.}
    \label{fig:fd}
    \end{center} \end{figure}

    Since we have tested the main ingredients of the model, we next turn to fragmentation reactions. Figure \ref{fig:aa} shows the charge distribution of $^{197}$Au+$^{197}$Au at 35 MeV/A with impact parameters ranging from 0 to 3.5 fm and t=1500 fm/c. The modified model seems to work better than the original one, especially in the light nuclei region, noticing that the original model gives too many protons (not displayed in the figure - out of scale)~\cite{papa2001}. At $Z=$20, the data show a jump due to the different detectors used~\cite{desequelles1998}. We would get a better agreement to the data with the calculations if we slightly shift the data yield up for $Z>$20~\cite{papa2001}. In fragmentation reactions, fluctuations are important since they could signal critical phenomena such as a liquid-gas phase transition. In order to test different physical observables, we calculated the fragment ($Z\ge4$) multiplicity ($N_{f4}$-Fig. \ref{fig:fd}a), the total charge bound in form of fragments ($Z_{b4}$-Fig. \ref{fig:fd}b), the fourth moment of the fragment charge distribution ($m_4$-Fig. \ref{fig:fd}c), and the charge asymmetry distribution (a$_{123}$-Fig. \ref{fig:fd}d) and compared to the experimental data~\cite{desequelles1998}. We took into account that the experimental detection efficiency of the total charge is at least 70\%, thus we modified the total charge bound in form of fragments in our calculations by a factor of 85\% (the average of 70\% and 100\%). For all the charge correlations shown, the modified model with the Heisenberg principle reproduced the data much better than the original CoMD model. After adding the Heisenberg principle, the light nuclei are less bound, thus the production of these particles is smaller while that of the heavier fragments becomes larger. This will lead to an increase in the total bounded charge and also in the fluctuation of the charge distribution.

    \begin{figure}[htbp] \begin{center}
        \includegraphics[width=\hsize]{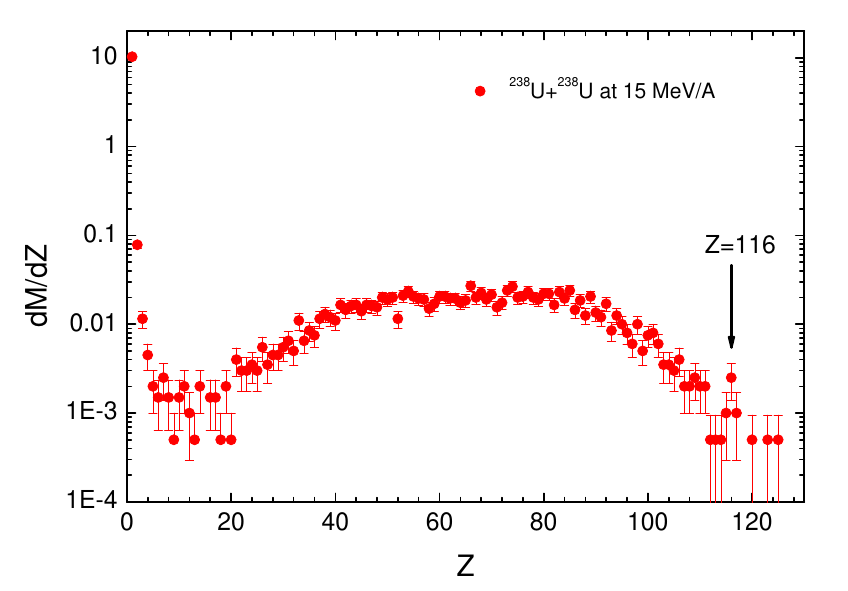}
        \caption{(Color online) Isotope distribution of $^{238}$U+$^{238}$U at 15 MeV/A with impact parameter of 0 fm and t=4000 fm/c using the CoMD$_H$ model.}
    \label{fig:uu}
    \end{center} \end{figure}

    \begin{figure}[htbp] \begin{center}
        \includegraphics[width=\hsize]{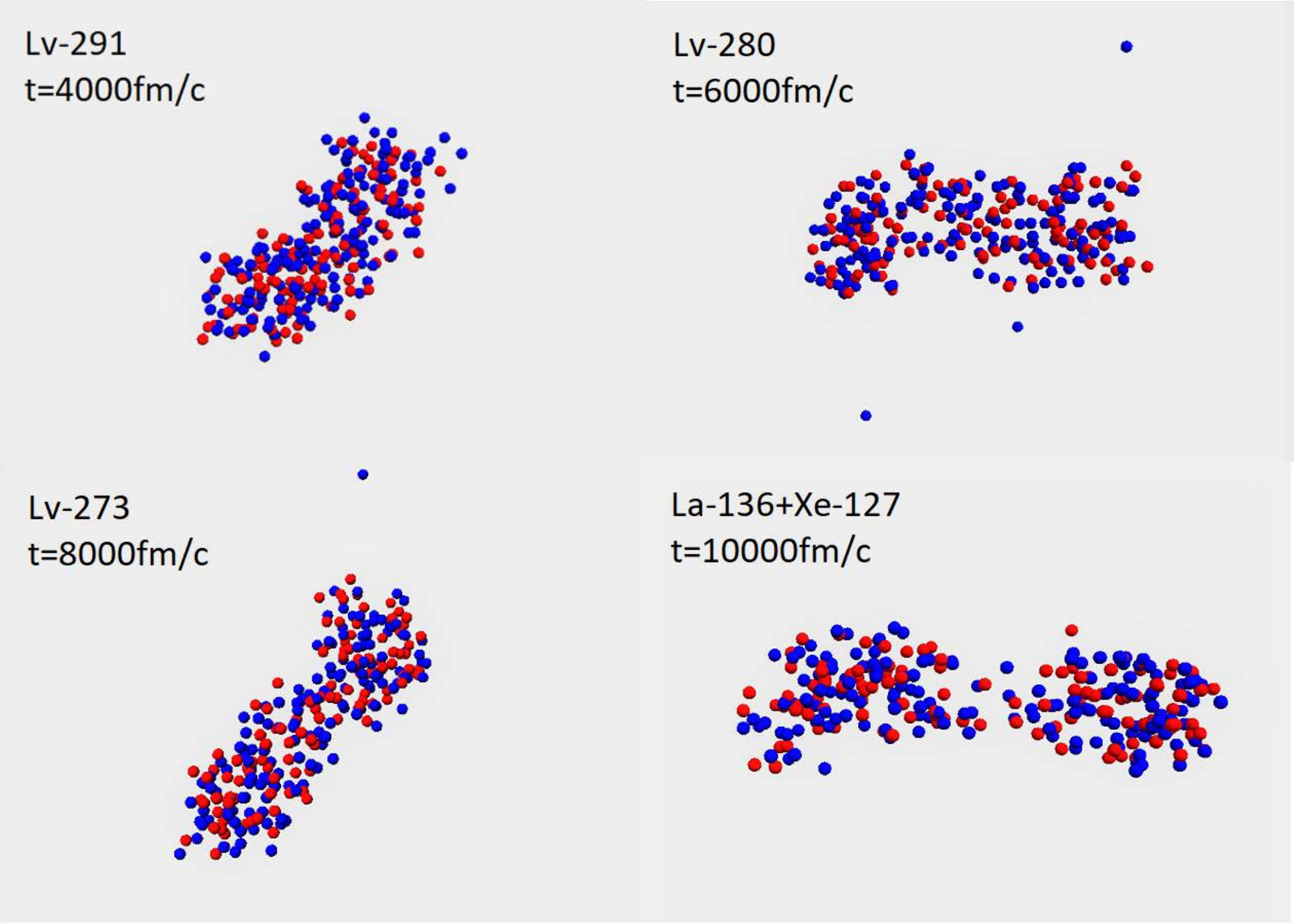}
        \caption{(Color online) Time evolution of Livermorium-291 using the CoMD$_H$ model. The red spheres represent protons, while the blue ones represent neutrons. Only the largest clusters are shown.}
    \label{fig:lv}
    \end{center} \end{figure}

    We performed calculations at lower beam energies to test the probability of superheavy formation in the collision of very heavy nuclei~\cite{maruyama2002,wuenschel2018}. Figure \ref{fig:uu} displays the charge distribution of the $^{238}$U+$^{238}$U at 15 MeV/A with zero impact parameter and t=4000 fm/c. The binding energies of some isotopes coming out from the reaction region are shown in Fig. \ref{fig:bech}. We see that all the binding energies of the nuclei from the collisions are smaller than those in the initialization step, since they are still excited after the 4000 fm/c evolution (in the model they cannot emit say gamma rays to decay to their ground states). The binding energies of deuterons and alpha particles of the modified model are smaller than those of the original one, while for heavy nuclei, they are more or less the same, consistent with the initialization step. In the superheavy region, a peak around Z=116 was found. We selected one nucleus with Z=116 and A=291, and let it evolve with time. One out of 500,000 events survived for a total time of another 6000 fm/c with only several neutrons emitted as shown in Fig. \ref{fig:lv}, then finally it underwent fission and broke into $^{136}$La+$^{127}$Xe and one alpha particle with 15.78 MeV kinetic energy (close to the value predicted by the nuclear density functional theory, which is around 14 MeV~\cite{staszczak2013,wuenschel2018}), and several protons and neutrons. This might be a different path to study super-heavy production~\cite{maruyama2002} and we will discuss more in detail in future works since a large calculation effort is required.

\section{Summary} \label{Sec:sum}

    In this paper, we have succeeded in modifying the CoMD model by introducing the Heisenberg principle. The binding energies of the light particles are more comparable to the experimental data in both initializations and collisions. The modified model works well with exotic nuclei. We have also performed some superheavy nuclear reactions with the new model. Though we got some events, more efforts, both numerically and theoretically, need to be done and we decided to postpone it until further refinements in the model are included such as the (iso)spin-(iso)spin forces.

\begin{acknowledgments}
    This work is partially supported by the National Natural Science Foundation of China under Contracts Nos. 11421505, 11305239, 11205230 and 11520101004, the Major State Basic Research Development Program in China under Contracts Nos.  2014CB845401 and 2013CB834405, the CAS Project Grant No. QYZDJSSW-SLH002, and also the Chinese Government Scholarship. This material is based upon work supported by the Department of Energy, National Nuclear Security Administration, under Award Number DE-NA0003841, the Center for Excellence in Nuclear Training And University-based Research (CENTAUR).
\end{acknowledgments}

%\bibliography{Giant Dipole Resonance in Proton Capture Reaction Using a Quantum Molecular Dynamics Model}
\bibliographystyle{apsrev4-1}
\providecommand{\noopsort}[1]{}\providecommand{\singleletter}[1]{#1}%
%

%\end{CJK*}
\end{document}